\documentclass[a4paper]{article}

\usepackage{19lomcon}        % Proceedings volume layout metrics
\usepackage{cite}             % Smart range citations
\usepackage{epsfig}           % Encapsulated PostScript figure inclusion
\usepackage{epstopdf}

\begin{document}

% MATH SYMBOLS
%
\def\cosech{\rm cosech}
\def\sech{\rm sech}
\def\coth{\rm coth}
\def\tanh{\rm tanh}
%fractions
\def\half{{1\over 2}}
\def\third{{1\over3}}
\def\fourth{{1\over4}}
\def\fifth{{1\over5}}
\def\sixth{{1\over6}}
\def\seventh{{1\over7}}
\def\eigth{{1\over8}}
\def\ninth{{1\over9}}
\def\tenth{{1\over10}}
\def\bN{\mathop{\bf N}}
\def\R{{\rm I\!R}}
\def\Eins{{\mathchoice {\rm 1\mskip-4mu l} {\rm 1\mskip-4mu l}
{\rm 1\mskip-4.5mu l} {\rm 1\mskip-5mu l}}}
\def\Z{{\mathchoice {\hbox{$\sf\textstyle Z\kern-0.4em Z$}}
{\hbox{$\sf\textstyle Z\kern-0.4em Z$}}
{\hbox{$\sf\scriptstyle Z\kern-0.3em Z$}}
{\hbox{$\sf\scriptscriptstyle Z\kern-0.2em Z$}}}}
\def\abs#1{\left| #1\right|}
\def\com#1#2{
        \left[#1, #2\right]}
\def\square{\kern1pt\vbox{\hrule height 1.2pt\hbox{\vrule width 1.2pt
   \hskip 3pt\vbox{\vskip 6pt}\hskip 3pt\vrule width 0.6pt}
   \hrule height 0.6pt}\kern1pt}
      \def\boxop{{\raise-.25ex\hbox{\square}}}
% \contract is a differential geometry contraction sign _|
\def\contract{\makebox[1.2em][c]{
        \mbox{\rule{.6em}{.01truein}\rule{.01truein}{.6em}}}}
\def\ltap{\ \raisebox{-.4ex}{\rlap{$\sim$}} \raisebox{.4ex}{$<$}\ }
\def\gtap{\ \raisebox{-.4ex}{\rlap{$\sim$}} \raisebox{.4ex}{$>$}\ }
\def\mn{{\mu\nu}}
\def\rs{{\rho\sigma}}
\newcommand{\Det}{{\rm Det}}
\def\Tr{{\rm Tr}\,}
\def\tr{{\rm tr}\,}
\def\sumij{\sum_{i<j}}
\def\e{\,{\rm e}}
%boldface vectors
\def\non{\nonumber\\}
\def\br{{\bf r}}
\def\bp{{\bf p}}
\def\bx{{\bf x}}
\def\by{{\bf y}}
\def\brhat{{\bf \hat r}}
\def\bv{{\bf v}}
\def\ba{{\bf a}}
\def\bE{{\bf E}}
\def\bB{{\bf B}}
\def\bA{{\bf A}}
%derivatives
\def\pa{\partial}
\def\dA{\partial^2}
\def\ddx{{d\over dx}}
\def\ddt{{d\over dt}}
\def\der#1#2{{d #1\over d#2}}
\def\lie{\hbox{\it \$}} % fancy L for the Lie derivative
\def\partder#1#2{{\partial #1\over\partial #2}}
\def\secder#1#2#3{{\partial^2 #1\over\partial #2 \partial #3}}
%
%equations
%\newcommand{\be}{\begin{equation}}
%\newcommand{\ee}{\end{equation}\noindent}
%\newcommand{\bear}{{\begin{eqnarray}}}
%\newcommand{\ear}{{\end{eqnarray}\noindent}}
%\newcommand{\benn}{\begin{enumerate}}
%\newcommand{\enn}{\end{enumerate}}
%\newcommand{\veject}{\vfill\eject}
%\newcommand{\ven}{\vfill\eject\noindent}
\def\be{\begin{equation}}
\def\ee{\end{equation}\noindent}
\def\bear{\begin{eqnarray}}
\def\ear{\end{eqnarray}\noindent}
\def\bec{\begin{equation}}
\def\eec{\end{equation}\noindent}
\def\bearc{\begin{eqnarray}}
\def\earc{\end{eqnarray}\noindent}
\def\benn{\begin{enumerate}}
\def\enn{\end{enumerate}}
\def\veject{\vfill\eject}
\def\ven{\vfill\eject\noindent}
%
%reference to equations
\def\eq#1{{eq. (\ref{#1})}}
\def\eqs#1#2{{eqs. (\ref{#1}) -- (\ref{#2})}}
%
%integrals
\def\totint{\int_{-\infty}^{\infty}}
\def\posint{\int_0^{\infty}}
\def\negint{\int_{-\infty}^0}
\def\pint{{\dps\int}{dp_i\over {(2\pi)}^d}}
%
% PHYS SYMBOLS
\newcommand{\GeV}{\mbox{GeV}}
\def\FFdual{F\cdot\tilde F}
\def\bra#1{\langle #1 |}
\def\ket#1{| #1 \rangle}
\def\braket#1#2{\langle {#1} \mid {#2} \rangle}
\def\vev#1{\langle #1 \rangle}
\def\rightvac{\mid 0\rangle}
\def\leftvac{\langle 0\mid}
\def\ihbar{{i\over\hbar}}
% dirac matrix stuff
\def\ge{\hbox{$\gamma_1$}}
\def\gz{\hbox{$\gamma_2$}}
\def\gd{\hbox{$\gamma_3$}}
\def\go{\hbox{$\gamma_1$}}
\def\gt{\hbox{\$\gamma_2$}}
\def\gth{\hbox{$\gamma_3$}} 
\def\gf{\hbox{$\gamma_5\;$}}
\def\slash#1{#1\!\!\!\raise.15ex\hbox {/}}
\newcommand{\slD}{\,\raise.15ex\hbox{$/$}\kern-.27em\hbox{$\!\!\!D$}}
\newcommand{\slpartial}{\raise.15ex\hbox{$/$}\kern-.57em\hbox{$\partial$}}
\newcommand{\PP}{\cal P}
\newcommand{\G}{{\cal G}}
\newcommand{\nc}{\newcommand}
\newcommand{\Fkala}{F_{\kappa\lambda}}
\newcommand{\Fkanu}{F_{\kappa\nu}}
\newcommand{\Flaka}{F_{\lambda\kappa}}
\newcommand{\Flamu}{F_{\lambda\mu}}
\newcommand{\Fmunu}{F_{\mu\nu}}
\newcommand{\Fnumu}{F_{\nu\mu}}
\newcommand{\Fnuka}{F_{\nu\kappa}}
\newcommand{\Fmuka}{F_{\mu\kappa}}
\newcommand{\Fkalamu}{F_{\kappa\lambda\mu}}
\newcommand{\Flamunu}{F_{\lambda\mu\nu}}
\newcommand{\Flanumu}{F_{\lambda\nu\mu}}
\newcommand{\Fkamula}{F_{\kappa\mu\lambda}}
\newcommand{\Fkanumu}{F_{\kappa\nu\mu}}
\newcommand{\Fmulaka}{F_{\mu\lambda\kappa}}
\newcommand{\Fmulanu}{F_{\mu\lambda\nu}}
\newcommand{\Fmunuka}{F_{\mu\nu\kappa}}
\newcommand{\Fkalamunu}{F_{\kappa\lambda\mu\nu}}
\newcommand{\Flakanumu}{F_{\lambda\kappa\nu\mu}}

\newcommand{\pb}{\bar{p}}
\newcommand{\ph}{\hat{p}}
\newcommand{\gb}{\bar{g}}
\newcommand{\gh}{\hat{g}}
\newcommand{\zb}{\bar{z}}
\newcommand{\zh}{\hat{z}}
\newcommand{\wh}{\hat w}
\newcommand{\wb}{\bar w}
\newcommand{\p}{p\!\!\!/~}
\newcommand{\pbdash}{\bar{p} \!\!\!/~}
\newcommand{\q}{q\!\!\!/~}
\newcommand{\B}{\beta \!\!\!/~}
\newcommand{\tb}{\bar{t}}

\nc{\spa}[3]{\left\langle#1\,#3\right\rangle}
\nc{\spb}[3]{\left[#1\,#3\right]}
\nc{\ksl}{\not{\hbox{\kern-2.3pt $k$}}}
\nc{\hf}{\textstyle{1\over2}}
\nc{\pol}{\varepsilon}
\nc{\tq}{{\tilde q}}
\nc{\esl}{\not{\hbox{\kern-2.3pt $\pol$}}}
\newcommand{\cL}{\cal L}
\newcommand{\D}{\cal D}
\newcommand{\Dhalf}{{D\over 2}}
\def\eps{\epsilon}
\def\epshalf{{\epsilon\over 2}}
\def\lag{( -\partial^2 + V)}
%worldline
\def\freeexp{{\rm e}^{-\int_0^Td\tau {1\over 4}\dot x^2}}
\def\kinb{{1\over 4}\dot x^2}
\def\kinf{{1\over 2}\psi\dot\psi}
\def\expk{{\rm exp}\biggl[\,\sum_{i<j=1}^4 G_{Bij}k_i\cdot k_j\biggr]}
\def\expp{{\rm exp}\biggl[\,\sum_{i<j=1}^4 G_{Bij}p_i\cdot p_j\biggr]}
\def\expshort{{\e}^{\half G_{Bij}k_i\cdot k_j}}
\def\expabb{{\e}^{(\cdot )}}
\def\epseps#1#2{\varepsilon_{#1}\cdot \varepsilon_{#2}}
\def\epsk#1#2{\varepsilon_{#1}\cdot k_{#2}}
\def\kk#1#2{k_{#1}\cdot k_{#2}}
\def\G#1#2{G_{B#1#2}}
\def\Gp#1#2{{\dot G_{B#1#2}}}
\def\GF#1#2{G_{F#1#2}}
\def\Dab{{(x_a-x_b)}}
\def\Dsq{{({(x_a-x_b)}^2)}}
\def\PITD{{(4\pi T)}^{-{D\over 2}}}
\def\4piTD{{(4\pi T)}^{-{D\over 2}}}
\def\4piT4{{(4\pi T)}^{-2}}
\def\TintmD{{\dps\int_{0}^{\infty}}{dT\over T}\,e^{-m^2T}
    {(4\pi T)}^{-{D\over 2}}}
\def\Tintm4{{\dps\int_{0}^{\infty}}{dT\over T}\,e^{-m^2T}
    {(4\pi T)}^{-2}}
\def\Tintm{{\dps\int_{0}^{\infty}}{dT\over T}\,e^{-m^2T}}
\def\Tint{{\dps\int_{0}^{\infty}}{dT\over T}}
\def\np{n_{+}}
\def\nm{n_{-}}
\def\Np{N_{+}}
\def\Nm{N_{-}}
\newcommand{\slG}{{{\dot G}\!\!\!\! \raise.15ex\hbox {/}}}
\newcommand{\Gd}{{\dot G}}
\newcommand{\Gund}{{\underline{\dot G}}}
\newcommand{\Gdd}{{\ddot G}}
\def\GBd12{{\dot G}_{B12}}
\def\Dx{\dps\int{\cal D}x}
\def\Dy{\dps\int{\cal D}y}
\def\Dpsi{\dps\int{\cal D}\psi}
\def\dint#1{\int\!\!\!\!\!\int\limits_{\!\!#1}}
\def\ddtau{{d\over d\tau}}
\def\ie{\hbox{$\textstyle{\int_1}$}}
\def\iz{\hbox{$\textstyle{\int_2}$}}
\def\id{\hbox{$\textstyle{\int_3}$}}
\def\ldop{\hbox{$\lbrace\mskip -4.5mu\mid$}}
\def\rdop{\hbox{$\mid\mskip -4.3mu\rbrace$}}
%
%VARIOUS
\newcommand{\1}{{\'\i}}
\newcommand{\no}{\noindent}
\def\non{\nonumber}
\def\dps{\displaystyle}
\def\sy{\scriptscriptstyle}
\def\sy{\scriptscriptstyle}

%%%%%%%%%%%%%%%%%%%%%%%%%%%%%%%%%%%%%%%%%%%%%%%%%%%%%%%%%%%%%%%%%%
% The preamble of the paper
%%%%%%%%%%%%%%%%%%%%%%%%%%%%%%%%%%%%%%%%%%%%%%%%%%%%%%%%%%%%%%%%%%

\title{MULTILOOP QED IN THE EULER-HEISENBERG APPROACH}
%%%%%%%%%%%%%%%%%%%%%%%%%%%%%%%%%%%%%%%%%%%%%%%%%%%%%%%%%%%%%%%%%%
% The preamble of the paper
%%%%%%%%%%%%%%%%%%%%%%%%%%%%%%%%%%%%%%%%%%%%%%%%%%%%%%%%%%%%%%%%%%
\author{Idrish Huet \email{idrish@ifm.umich.mx}}
\affiliation{Facultad de Ciencias en F\'isica y Matem\'aticas, Universidad Aut\'onoma de Chiapas,
 Ciudad Universitaria, Tuxtla Guti\'errez 29050, Mexico}
\author{Michel Rausch de Traubenberg \email{Michel.Rausch@iphc.cnrs.fr}}
\affiliation{IPHC-DRS, UdS, IN2P3,
23 rue du Loess, F-67037 Strasbourg Cedex, France}
\author{\underline{Christian Schubert} \email{christian.schubert@umich.mx}}
\affiliation{Instituto de F{{\'\i}}sica y Matem\'aticas, Universidad Michoacana de San Nicol\'as de Hidalgo,
Apdo. Postal 2-82, C.P. 58040, Morelia, Michoacan, Mexico}

% You may repeat \author and \affiliation as many times as necessary!

\date{}
% Print it out!
\maketitle

\begin{abstract}
I summarize what is known about the Euler-Heisenberg Lagrangian and its multiloop corrections for scalar and spinor QED, in various types of constant fields, and in various dimensions. Particular attention is given to the asymptotic properties of the weak-field expansion of the Lagrangian, which via Borel summation is related to Schwinger pair-creation, and the status of the "exponentiation conjecture" for the imaginary part of the Euler-Heisenberg to all loop orders. 

\end{abstract}

%\section{Euler-Heisenberg Lagrangian and photon amplitudes}

In 1936 Heisenberg and Euler obtained their well-known
effective Lagrangian in a constant field (`` Euler-Heisenberg Lagrangian'' = `EHL')
in the form

\bear
{\cal L}^{(1)}(a,b)&=& - {1\over 8\pi^2}
\int_0^{\infty}{dT\over T^3}
\,\e^{-m^2T}
\biggl\lbrack
{(eaT)(ebT)\over {\rm tanh}(eaT)\tan(ebT)} 
%\nonumber\\&&\hspace{70pt}
- {e^2\over 3}(a^2-b^2)T^2 -1
\biggr\rbrack
\, .
\label{ehspin}
\nonumber\\
\ear
Here  $a,b$  are the two
invariants of the Maxwell field, 
related to $\bf E$, $\bf B$ by $a^2-b^2 = B^2-E^2,\quad ab = {\bf E}\cdot {\bf B}$.
The superscript $(1)$ stands for one-loop. 

The EHL describes the effective interaction between photons due to the presence of virtual electron-positron pairs in the vacuum, leading to light-by-light scattering, vacuum birefringence in a magnetic field,
Schwinger pair creation in an electric field, and many other effects of actual interest for high-energy
and laser physics. 
Fourier transformation of the EHL yields the $N$ - photon amplitudes in the low energy limit, where all photon energies are small compared to the electron mass, $\omega_i\ll m$.
Diagrammatically, this corresponds to Fig. \ref{fig-EHL1loop}:

\begin{figure}[htbp]
\begin{center}
\includegraphics[scale=.5]{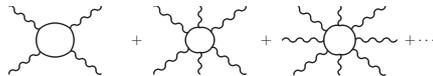}
%\caption{{\bf }}
\end{center}
\caption{Sum of diagrams equivalent to the one-loop EHL.}
\label{fig-EHL1loop}
\end{figure}
%
%In \cite{56} it was shown how to construct these amplitudes explicitly 
%from the weak field expansion coefficients  $c_{kl}$,  defined by
%
%\bear
%{\cal L}^{(1)} (a,b) = \sum_{k,l} c_{kl}\, a^{2k}b^{2l} \, .
%\ear
%In particularly, there it was shown that,  
%for each $ N$ and each given helicity assignment, the dependence on the momentum and polarization vectors can be absorbed into a  single invariant $ \chi_N$.
%

%\section{Imaginary part and Sauter-Schwinger pair creation}

If the field has an electric component ($ b \ne 0$) then there are poles on the 
integration contour at $ ebT =  k\pi$ which create an imaginary part.
For the purely electric case one gets Schwinger's 1951 formula,

\begin{eqnarray}
{\rm Im} {\cal L}^{(1)}(E) &=&  \frac{m^4}{8\pi^3}
\beta^2\, \sum_{k=1}^\infty \frac{1}{k^2}
\,\exp\left[-\frac{\pi k}{\beta}\right]
%\non\\
%{\rm Im}{\cal L}_{\rm scal}^{(1)}(E) 
%&=&
%-\frac{m^4}{16\pi^3}
%\beta^2\, \sum_{k=1}^\infty \frac{(-1)^{k}}{k^2}
%\,\exp\left[-\frac{\pi k}{\beta}\right]
\label{schwinger}
\end{eqnarray}
($\beta = eE/m^2$). 
In the following we will consider the weak-field limit  $\beta \ll 1$, where only the leading $k=1$ is relevant. 
Note that $ {\rm Im}{\cal L}^{(1)}(E)$ depends on $ E$  non-perturbatively (nonanalytically), which is consistent with Sauter's 
interpretation of pair creation as  vacuum tunneling.
%
% (Fig. \ref{fig-pairtunnel}). 
%
%\begin{figure}[h]
%%\begin{picture}(0,0)(6000,10000)
%\centerline{\centering
%\includegraphics[scale=.7]{fig-pairtunnel.eps}
%}
%%\end{picture}
%\caption{Pair creation by an external field.}
%\label{fig-pairtunnel}
%\end{figure}

\begin{figure}[ht]
\centerline{\hspace{30pt}\includegraphics[scale=.38]{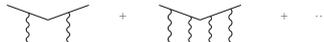}}
%   \includegraphics{fig1.eps}
%\end{figure}

\no
\caption{``Cut'' diagrams describing Schwinger pair creation.}
\label{fig-schwinger_tree_diags}
\end{figure}

\no
This connection between the effective action and the pair creation rate is based on the  Optical Theorem, which relates
the imaginary part of the diagrams shown in Fig. \ref{fig-EHL1loop} to the `` cut diagrams '' shown in Fig. \ref{fig-schwinger_tree_diags}.

\no
However, the latter individually all vanish for a constant field, which can emit only zero-energy photons.
Thus for a constant field we cannot use dispersion relations for individual diagrams;
what counts is the asymptotic behaviour of the diagrams for a large number of photons. 
The appropriate generalization is a Borel dispersion relation. This works in the following way \cite{37}:
define the  weak field expansion coefficients of the purely electric EHL by

\bear
{\cal L}^{(1)}(E) = \sum_{n=2}^{\infty} c(n) \Bigl(\frac{eE}{m^2}\Bigr)^{2n} \, .
\ear
It can be shown that their leading large - $n$ behavior is

\bear
c(n)\,\, {\stackrel{n\to \infty}{\sim}} \,\,c_{\infty} \Gamma[2n - 2] 
\label{leading}
\ear
with a constant $c_{\infty}$. 
The Borel dispersion relation relates this leading behavior to the leading weak-field behavior
of the imaginary part of the Lagrangian:

\bear
{\rm Im}{\cal L}^{(1)}(E) \, \, {\stackrel{\beta\to 0}{\sim}} \,\,
c_{\infty}\,\e^{-\frac{\pi m^2}{eE}} \, .
\ear
Thus we have rederived the leading Schwinger exponential of (\ref{schwinger}) in a way
that turns out to be very useful for higher-loop considerations. 

\begin{figure}[h]
%\begin{picture}(0,0)(6000,10000)
\centerline{\centering
\includegraphics[scale=.37]{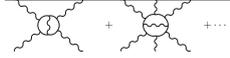}
}
\caption{Feynman diagrams corresponding to the 2-loop EHL.}
\label{fig-2loopEHL}
\end{figure}

\begin{figure}[h]
\centerline{\centering
\includegraphics[scale=.2]{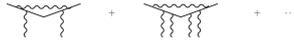}
}
\caption{Feynman diagrams contributing to 2-loop Schwinger pair creation.}
\label{fig-schwinger_tree_adorned_diags}
\end{figure}

The two-loop (one-photon exchange) correction to the EHL corresponds to the set of diagrams shown in Fig. 
\ref{fig-2loopEHL} (there is also a one-particle reducible contribution \cite{giekar}, but for our present purposes it can be discarded).
The corresponding corrections to the tree-level pair creation diagrams of Fig. \ref{fig-schwinger_tree_diags} are shown in Fig. 
\ref{fig-schwinger_tree_adorned_diags}.

These two-loop diagrams lead to rather intractable two-parameter integrals
\cite{ritusspin,lebrit}. 
However, the imaginary part $ {\rm Im}{\cal L}^{(2)}(E)$ 
when taken in the weak-field limit becomes a simple addition to $ {\rm Im}{\cal L}^{(1)}(E)$ \cite{lebrit}:
 
% \vspace{-10pt}
 \bear
{\rm Im} {\cal L}^{(1)} (E) +
{\rm Im}{\cal L}^{(2)} (E) 
\,\,\,\, {\stackrel{\beta\to 0}{\sim}} \,\,\,\,
 \frac{m^4\beta^2}{8\pi^3}
\bigl(1+\alpha\pi\bigr)
\,{\rm e}^{-{\pi\over\beta}}
\, .
\label{Im1plus2}
\ear

\no
In \cite{lebrit}, Lebedev and Ritus further noted that, if one assumes that higher orders 
in this limit will lead to exponentiation,

 \bear
{\rm Im} {\cal L}^{(1)} (E) +
{\rm Im}{\cal L}^{(2)} (E) 
%+ {\rm Im}{\cal L}^{(3)} (E) 
+ \ldots
\,\,\,\, {\stackrel{\beta\to 0}{\sim}} \,\,\,\,
 \frac{m^4\beta^2}{8\pi^3}
\,{\rm exp}\Bigl[ -{\pi\over\beta}+\alpha\pi \Bigr]
= {\rm Im}{\cal L}^{(1)}(E)\,\,{\rm e}^{\alpha\pi}
\ear
then the result could be interpreted in the tunneling picture as the  corrections to the
Schwinger pair creation rate due to the pair being created with a negative Coulomb
interaction energy:

\bear
m(E) \approx m + \delta m(E), \quad \delta m(E) =  - \frac{\alpha}{2}\frac{eE}{m} 
\ear
where $\delta m (E)$ is just the ``Ritus mass shift'', originally derived from the
crossed process of one-loop electron propagation in the field \cite{ritusmass}. 

% (Fig. \ref{crossing}). 
%
%\vspace{-5pt}
%
%\begin{figure}[h]
%%\begin{minipage}[b]{0.5\linewidth}
%\centering
%\hspace{10pt}\includegraphics[scale=.5]{fig-uncrossed.eps}
%\raisebox{3.1 em} {$\hspace{20pt}\Longleftrightarrow\hspace{20pt}$}
%\includegraphics[scale=.4]{fig-crossed.eps}
%\hspace{20pt}
%\label{fig4point}
%%\end{minipage}
%%\hspace{0cm}
%%\begin{minipage}[b]{0.5\linewidth}
%%\centering
%%\includegraphics[scale=.8]{fighexagon.eps}
%\caption{Crossing relation between pair creation and electron propagation in the field.}
%\label{fig-crossing}
%%\end{minipage}
%\end{figure}
%%\hspace{70pt}  Pair creation by the field  
%%\hspace{30pt}  Electron propagation in the field

For  scalar QED the corresponding exponentiation had been conjectured already
two years earlier by Affleck, Alvarez and Manton \cite{afalma}:

\bear
\sum_{l=1}^{\infty}{\rm Im}{\cal L}^{(l)}_{\rm scal}(E)
&{\stackrel{\beta\to 0}{\sim}}&
 -\frac{m^4\beta^2}{16\pi^3}
\,{\rm exp}\Bigl[ -{\pi\over\beta}+\alpha\pi \Bigr] =  {\rm Im}{\cal L}^{(1)}_{\rm scal}(E)\,\,{\rm e}^{\alpha\pi} \, .
\label{ImLallloop}
\ear
However, they arrived at this conjecture in a very different way, namely
using Feynman's  worldline path integral formalism in a semi-classical
approximation. 

Thus there is much that speaks for the exponentiation conjecture. However, it would be a very surprising result, since it leads to 
the analytic factor $ {\rm e}^{\alpha\pi}$! 
This is counter-intuitive, since the growth in the number of diagrams caused by the insertion of an increasing number of photons into an electron loop 
would lead one to expect a vanishing radius of convergence in $\alpha$. 

Using Borel analysis, this factor can be transferred from the imaginary part of the effective Lagrangian to the large - $N$ limit of the $N$ - photon amplitudes \cite{60}: 

\vspace{-10pt}
\bear
{\rm lim}_{N\to\infty}
\frac
{\Gamma^{({\rm all-loop})}[k_1,\varepsilon_1;\ldots;k_N,\varepsilon_N]}
 {\Gamma^{(1)}[k_1,\varepsilon_1;\ldots;k_N,\varepsilon_N]}
=
\e^{\alpha\pi}
\, 
\label{conj}
\ear
%Here an essential ingredient is the above-mentioned fact that, independently of the loop order, the complete dependence of the $N$ - photon amplitudes
%for a fixed helicity assignment can be absorbed into a single invariant. 

%\section{The EHL in 1+1 dimensional QED}

The exponentiation conjecture has been verified at the two-loop order by explicit computation in both scalar and spinor QED. 
A three-loop check in $ D=4$ seems presently out of reach. 
In 2005, Krasnansky \cite{krasnansky} found the following explicit formula for the two-loop EHL
in scalar QED in 1+1 dimensions:

\vspace{-10pt}
\bear
{\cal L}_{\rm scal}^{(2)(2D)}(\kappa)
&=&
-\frac{e^2}{32\pi^2}\left(
\xi^2_{2D} 
-4\kappa \xi_{2D}'\right) ,
\ear
where $\xi_{2D}= -\Bigl(\psi(\kappa+\half)-\ln (\kappa)\Bigr)$,
$\psi(x)=\frac{\Gamma^\prime(x)}{\Gamma(x)}$, $\kappa =  \frac{m^2}{2ef}$, 
$f^2=\fourth F_{\mu\nu}F^{\mu\nu}$.

This is simpler than in four dimensions, but still non-trivial (in fact very similar in structure to the EHL
in four dimensions for a self-dual field \cite{51}), which suggests to use 1+1 dimensional QED as a toy model for 
testing the exponentiation conjecture.  
In \cite{81} we used the method of \cite{afalma} to generalize the exponentiation conjecture to the 2D case, obtaining

\vspace{-10pt}
\bear
{\rm Im}{\cal L}^{(all-loop)}_{2D}
\sim
\e^{-\frac{m^2\pi}{eE} + \tilde\alpha \pi^2  \kappa^2}
\label{2Dexp}
\ear
where $\tilde\alpha = \frac{2e^2}{\pi m^2}$, and obtained
the one- and two-loop $2D$ spinor QED EHL
explicitly in terms of the function
%\bear
%{\cal L}^{(1)}_{2D}(\kappa ) &=& -{m^2\over 4\pi} {1\over\kappa}
%\Bigl[{\rm ln}\Gamma(\kappa) - \kappa(\ln \kappa -1) +
%\half \ln \bigl({\kappa\over 2\pi}\bigr)\Bigr] \, ,
%\nonumber\\
%{\cal L}^{(2)}_{2D}(\kappa) &=& {m^2\over 4\pi}\frac{\tilde\alpha}{4}
%\Bigl[ \tilde\psi(\kappa) + \kappa \tilde\psi'(\kappa)
%+\ln(\lambda_0 m^2) + \gamma + 2 \Bigr] \, ,
%\nonumber
%\ear \no
$\tilde\psi (x) \equiv \psi(x) - \ln x + {1\over 2x}$.
This allowed us to obtain formulas for $ c^{(1,2)}(n)$ 
in terms of the Bernoulli numbers,
%\bear
%c^{(1)}(n) &=& (-1)^{n+1} \frac{B_{2n}}{4n(2n-1)}\, ,\nonumber\\
%c^{(2)}(n) &=& (-1)^{n+1} \frac{\tilde\alpha}{8}\frac{2n-1}{2n}B_{2n} \, .\nonumber
%\ear
and made it possible to analytically verify
(\ref{2Dexp}) at the two-loop level:

\vspace{-10pt}
\bear
\lim_{n\to\infty}  {c^{(2)}(n)\over c^{(1)}(n+1)} 
&=&
\tilde\alpha \pi^2 \, .
\ear
%(the relative shift in the argument of the coefficients is due to the fact that, unlike the four-dimensional case, 
%in two dimensions the term involving $\tilde\alpha$ in the exponent in (\ref{2Dexp}) also involves the external field).
%The convergence of $ c^{(2)}(n)$ to the asymptotic prediction is rather fast (Fig. \ref{fig-ratio21}):
%
%\begin{figure}[h]
%%\begin{picture}(0,0)(6000,10000)
%\centerline{\centering
%\includegraphics[scale=.55]{fig-ratio21.eps}
%}
%\caption{Convergence of the two-loop coefficients to the asymptotic prediction.}
%\label{fig-ratio21}
%\end{figure}
%\bigskip

%\section{The three-loop EHL in 1 + 1 dimensional QED}

\begin{figure}[h]
%\begin{picture}(0,0)(6000,10000)
\centerline{\centering
\includegraphics[scale=.2]{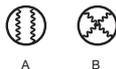}
}
\vspace{-9pt}
\caption{Single electron-loop contributions to the three-loop EHL.}
\label{fig-AB}
\end{figure}

At the three-loop level, only the two diagrams with a single electron loop, Fig. 
\ref{fig-AB}, turn out to be
relevant for the exponentiation conjecture. 

\begin{figure}[h]
%\begin{picture}(0,0)(6000,10000)
\centerline{\centering
\includegraphics[scale=.2]{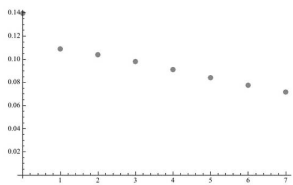}
}
\caption{Plot of the three-loop coefficients against the asymptotic prediction.}
\label{fig-ratio31}
\end{figure}

%In $D=2$, the EHL is already UV finite at the three-loop level. There are spurious IR divergences,
%but they can be removed by choosing the ``traceless gauge'' $\xi = -2$. 
In \cite{121} the authors
developed a method for computing the weak-field expansion coefficients for these diagrams,
and obtained the first seven coefficients explicitly; one more was computed by E. Panzer. 
Plotting these eight coefficients, one gets Fig. \ref{fig-ratio31},
where the normalization is such that the coefficients would have to converge to unity for the conjecture
to hold at the three-loop level, which appears unlikely.

\bigskip

\noindent
To summarize:

\begin{itemize}

\item
The higher-loop corrections to the one-loop Euler-Heisenberg
Lagrangian should provide valuable information on the
asymptotic properties of QED.

\item
The exponentiation conjecture implies two very non-obvious properties of the
asymptotic growth of the weak-field expansion coefficients:
(i) At any fixed loop order, their leading asymptotic growth is always the same (which already at the two-loop level
is true only after renormalization!).
(ii) The coefficients of these leading terms exponentiate in $\alpha$.

\item
The exponentiation in $\alpha$ seems not to be true (at least not for
$D = 2$), but analyticity in $\alpha$ remains a strong possibility.

\end{itemize}

%%%%%%%%%%%%%%%%%%%%%%%%%%%%%%%%%%%%%%%%%%%%%%%%%%%%%%%%%%%%%%%%%%
% References
%%%%%%%%%%%%%%%%%%%%%%%%%%%%%%%%%%%%%%%%%%%%%%%%%%%%%%%%%%%%%%%%%%


\begin{thebibliography}{99}

\bibitem{37} 
G.V. Dunne, C. Schubert, 
%``Two-loop Euler-Heisenberg QED pair-production rate'', 
Nucl. Phys. {\bf B 564} (2000) 59.
%hep-th/9907190. 

\bibitem{giekar}
H. Gies, F. Karbstein,
%``An Addendum to the Heisenberg-Euler effective action beyond one loop'',
JHEP {\bf 1703} (2017) 108. 
%arXiv:1612.07251 [hep-th].  

\bibitem{ritusspin}
V. I. Ritus, 
%``Lagrangian of an intense electromagnetic field and quantum electrodynamics at short distances'', 
Zh. Eksp. Teor. Fiz {\bf 69} (1975) 1517 [Sov. Phys. JETP {\bf 42} (1975) 774].

\bibitem{lebrit}
S. L. Lebedev, V. I. Ritus, 
%``Virial representation of the imaginary part of the Lagrangian function of an electromagnetic field'', 
Zh. Eksp. Teor. Fiz. {\bf 86} (1984) 408 [JETP {\bf 59} (1984) 237].

\bibitem{ritusmass} 
V.I. Ritus,
% ``Method of eigenfunctions and mass operator in quantum electrodynamics of a constant field'', 
Zh. Eksp. Teor. Fiz. {\bf 75} (1978) 1560 [JETP {\bf 48} (1978) 788].

\bibitem{afalma}
I.K. Affleck, O. Alvarez, N.S. Manton, 
%``Pair Production At Strong Coupling In Weak External Fields'',% 
Nucl. Phys. {\bf B 197} (1982) 509.

\bibitem{60}
G.V. Dunne, C. Schubert,
%`` Multiloop information from the QED effective lagrangian'', 
%IX Mexican Workshop on Particles and Fields, Colima 2005,  eds Paolo Amore et al,
J. Phys.: Conf. Ser. {\bf 37} (2006) 59. 
%hep-th/0409021.

\bibitem{krasnansky}
M. Krasnansky,
% ``Two-loop vacuum diagrams in background field and Heisenberg-Euler effective action in different dimension'',  
Int. J. Mod. Phys. {\bf A 23}, 5201 (2008).
%hep-th/0607230.

\bibitem{51}
G. V. Dunne, C. Schubert, 
%``Two-loop self-dual Euler-Heisenberg Lagrangians. I: Real part and helicity amplitudes'',
JHEP {\bf 0208} (2002) 053.
% hep-th/0205004. 

\bibitem{81}
I. Huet, D.G.C. McKeon, C. Schubert,
%``Euler-Heisenberg lagrangians and asymptotic analysis in 1+1 QED. Part I: two-loop",
JHEP {\bf 1012} (2010) 036. 
%\arXiv:1010.5315 [hep-th].
  
\bibitem{121}
I. Huet, M. Rausch de Traubenberg, C. Schubert,
%``Three-loop Euler-Heisenberg Lagrangian in 1+1 QED, part 1: single fermion-loop part'',
JHEP {\bf 1903} (2019) 167.
%arXiv:1812.08380[hep-th].

\end{thebibliography}
\end{document}